\documentclass[twocolumn]{autart}
\usepackage{color}
\usepackage[dvips]{epsfig}
\usepackage{amsmath}
\usepackage{amssymb}
\usepackage{subfigure}
\usepackage{txfonts}
\usepackage{mathtools}
\newtheorem{theorem}{Theorem}{}
\newtheorem{assumption}{Assumption}{}
\newtheorem{remark}{Remark}{}
\newtheorem{lemma}{Lemma}{}
\newtheorem{definition}{Definition}{}
\newtheorem{proposition}{Proposition}{}
\newcommand{\reg}{$^{\tiny\mbox{\textregistered}}\ $}
\usepackage{picinpar}

\begin{document}

\begin{frontmatter}

\title{Output Feedback Based Event-Triggered Sliding Mode Control for Delta Operator Systems\thanksref{footnoteinfo}}
\thanks[footnoteinfo]{This paper was not presented at any IFAC meeting. Corresponding author is Bijnan Bandyopadhyay. Tel. +91 22 2576 7889. Fax +91 22 2570 0057.}

\author[I]{Kiran Kumari}\ead{kiran@sc.iitb.ac.in},
\author[I]{Bijnan Bandyopadhyay}\ead{bijnan@sc.iitb.ac.in},
\author[II]{Kyung-Soo Kim}\ead{kyungsookim@kaist.ac.kr},
\author[III]{Hyungbo Shim}\ead{hshim@snu.ac.kr}

\address[I]{Systems and Control Engineering, Indian Institute of Technology Bombay, India,}
\address[II]{Department of Mechanical Engineering, KAIST, Daejeon, 305-701, Republic of Korea,}
\address[III]{ASRI, Department of Electrical and Computer Engineering, Seoul National University, Seoul, Korea}


\begin{abstract}
In this paper, we present an output feedback based design of event-triggered sliding mode control for delta operator systems. For discrete-time systems, multi-rate output sampling based state estimation technique is very useful if the output information is available. But at high sampling rates, the discrete-time representation of the system using shift operator becomes numerically ill-conditioned and as a result, the observability matrix becomes singular as the sampling period tends to zero. Here, a new formulation of multi-rate state estimation (MRSE) for a small sampling period is presented. We first propose a new observability matrix and then discuss its relationship with the observability matrix defined in the conventional sense. For the delta operator system with matched uncertainty, we have presented the design of MRSE based sliding mode control (SMC). Additionally, to make the control efficient in terms of resource utilization, MRSE based event-triggered SMC is proposed. The absence of Zeno phenomenon is guaranteed as the control input is inherently discrete in nature. Finally, the effectiveness of the proposed method is illustrated through numerical simulations, considering a ball and beam system and a general linear system as a numerical example.
\end{abstract}
\end{frontmatter}

\section{Introduction}
\vspace{-0.10in}

The use of digital computers for the implementation of controllers has widely increased in recent times. This is due to the availability of sophisticated and reliable digital computing platforms. For this purpose, the systems are generally analyzed in the discrete-time domain and conventional representation of the discrete-time system is done using a forward shift operator. To avoid any loss of information, typically a very high sampling frequency is used. But as the sampling frequency is increased, the system becomes numerically ill-conditioned because the input matrix of such a system tends to zero. To address this problem, Goodwin $\&$ Middleton \cite{delta} have defined the delta operator as
\begin{align*}
\delta x\left(t\right)=\begin{cases} \frac{\mathrm{d}x\left(t\right)}{\mathrm{d}t} \quad &\tau = 0\\
\frac{x(t+\tau)-x\left(t\right)}{\tau} \quad &\tau\neq 0
\end{cases}.
\end{align*}
The delta operator acts as a bridge between the continuous time and the discrete-time representations of the system and thus allows taking advantage of both the representations simultaneously \cite{high,rapp,uniapp}. Typically, there are uncertainties in modeling the system and often external disturbances are also active during the system operation. To deal with such situations robust controllers are designed, for example, high gain control, $\mathcal{H}_\infty$ control and SMC. In particular, SMC has received much attention because of its capability to reject matched disturbances \cite{edward,utkin}. As the discrete-time representation of systems is extensively used in various applications, the design of discrete-time sliding mode (DTSM) control for such systems has become necessary \cite{furuta,var}. As a consequence, many researchers have proposed different reaching laws \cite{barto,stab}. Over the last two decades, the modeling of different systems and the design of $\mathcal{H}_\infty$ control using the delta operator has increased widely \cite{Hinf,ext} but the dedicated design of control for the delta operator system has been dealt only in a few papers \cite{sohro}. Some work on the SMC design for the delta operator system is presented in \cite{novel}.

\quad However, generally, only the output measurements of a system are available, and full state information is needed to design state feedback control. Then the unmeasured states of the system are often estimated using observers. One such discrete estimating technique is multi-rate output feedback (MROF). For the last several decades, MROF technique has been extensively studied in the literature due to its ability to realize any full state feedback control by partial measurements without relying on the state estimation dynamics \cite{plant}. MROF can be thought of as a static state estimation technique utilizing the fast sub-samples \cite{werner}. In this paper, we call MROF based state estimation technique as MRSE. In spite of the notable advantages of MRSE, it can suffer from noise sensitivity and numerical inaccuracy for the very fast sampling. In \cite{jana2}, the multi-rate output sampling framework has been adopted for realizing the full state vector and applied to DTSM control. 

\quad On the other hand, in practice, control is implemented periodically which results in a large number of control updating instants for very high sampling frequency. Therefore, to make judicious use of available resources, it is advantageous to apply control only when it is needed and for this, Tabuada has developed a new aperiodic implementation technique known as \textit{event-triggering} \cite{tabuada,task}. In this strategy, control is updated when a certain rule is violated. The initial works on event-triggering have dealt with the system without uncertainties and only a few papers have dealt with uncertain systems \cite{feed,meng,sepa}. Recently, SMC has been incorporated with this approach because of its robustness property \cite{event5,euk,pract,perevent}. The event-triggered SMC for linear systems was presented in \cite{event1}. Later, it was also extended to discrete-time systems \cite{event2}. As discussed above, discrete-time systems become numerically ill-conditioned for high sampling frequency. To overcome this limitation and get the advantage of event-triggering, event-triggered control was first designed for the delta operator system in \cite{dek}. Moreover, in practical systems like mechanical systems and electrical systems, the sensor has a higher sampling frequency than the actuator and mostly only the output measurements are available. So the use of MRSE for such system is also desirable.

\quad The main contributions of the paper can be framed as follows: We are using past output samples to estimate the unmeasured states of the system. The numerical inaccuracy is the main focus of this paper which has not been addressed well in literature. In particular, when MRSE is realized for the closed-loop system, the reconstruction of the full state may not be feasible in practice due to the numerical inaccuracy. Therefore, in this paper, a new observability matrix for MRSE technique is proposed. This can be related to the observability matrix of the delta operator system. The newly proposed observability matrix approaches its continuous time counterpart as the sampling period tends to zero and it leads to a new formulation of MRSE technique to improve the numerical accuracy. In the case of the small sampling period, the control needs to be applied very frequently which increases the control computational burden. Along with this, the communication between the sensor and the controller also increases. Therefore, to reduce this, it is appealing to study event-triggered control for the delta operator systems. Moreover, to deal with uncertainties in systems, SMC is incorporated in the control design and an output feedback based event-triggered SMC is proposed for delta operator systems. As a consequence, the boundedness of the system trajectories with a reduced number of control updating instants is shown even in the presence of matched uncertainties.
 
\quad This paper is organized as follows: Section \ref{pre} discusses the preliminaries required to develop the theoretical results of the paper. Section \ref{mr1} \& \ref{mr2} establishes the main results and Section \ref{ne} presents the simulation results. Lastly, the concluding remarks are given in Section \ref{con}.

\textbf{Notation:}
We denote by $\mathbb{R}$ and $\mathbb{R}^n$ the set of real numbers and the set of $n$-dimensional real vectors respectively. $\mathbb{Z}$ denotes the set of integers and $\mathbb{Z}_{\geq 0}$ denotes the set of non-negative integers. The notation $\| \cdot\|$ denotes the Euclidean norm of a vector and $|\cdot |$ denotes the absolute value of a scalar. $\inf\left(\cdot\right)$ denotes the infimum of the argument. $\operatorname{sgn}\left(\cdot\right)$ denotes the signum operator. $blockdiag\{a,b,\cdots,c\}$ represents the diagonal matrix. $\lambda_{\min}\left(\cdot\right)\left(\lambda_{\max}\left(\cdot\right)\right)$ is the minimum(maximum) eigenvalue of the argument matrix.
\vspace{-0.05in}

\section{Preliminaries} \label{pre}
\vspace{-0.10in}

This section presents a brief introduction to the delta operator systems followed by the design of SMC for the same. It also discusses MRSE for the discrete-time system. 
\vspace{-0.05in}

\subsection{Delta operator systems}
\vspace{-0.10in}

Consider the following LTI continuous-time system
\vspace{-0.4in}

\begin{align}
\begin{split}
\label{sys1}
\dot{\xi}\left(t\right)&=A\xi\left(t\right)+B\left(u\left(t\right)+d\left(t\right)\right),\\
y\left(t\right)&=C\xi\left(t\right),
\end{split}
\end{align}
\vspace{-0.35in}

where $\xi \in \mathbb{R}^n$ represents the state vector, $y \in \mathbb{R}^p$ is the output of the system and $u \in \mathbb{R}$ represents the control input to the system. $d\in \mathbb{R}$ is the external disturbance acting on the system and the initial condition of the states is denoted as $\xi(0)=\xi_0$. All the matrices of the system are of appropriate dimensions.
\vspace{-0.10in}

\quad We have made the following assumptions about the system and they are assumed to hold throughout the paper. 
\vspace{-0.09in}

\begin{assumption} \label{assum1}
The disturbance $d$ is assumed to be bounded i.e., $\left| d(t)\right| <d_0$ for all $t\geq 0$. 
\end{assumption}
\vspace{-0.09in}

\begin{assumption} \label{assum2}
The pair $\left(A,B\right)$ is controllable and the pair $\left(A,C\right)$ is observable.
\end{assumption}
\vspace{-0.09in}

\begin{assumption} \label{assum3}
The disturbance is slowly varying with time $t$ and is constant in small sampling intervals. 
\end{assumption}
\vspace{-0.09in}

\begin{remark}
Note that, independent of the above assumption if the discretized system has the relative degree $1$ and it is minimum phase, an equivalent matched and bounded disturbance can be obtained by the method presented in \cite{shimd}. Therefore, if the disturbance for the delta operator system is found by the same method, the rest of the derivation and results of this paper will remain unchanged. Moreover, the discrete-time system has a relative degree $1$ for almost all sampling times no matter what relative degree the continuous-time system has \cite{shimd}. 
\end{remark}
\vspace{-0.05in}

\begin{remark}
Though we have used Assumption \ref{assum3} for the proposed method, we will demonstrate in Example $1$ that the proposed strategy is still effective for slowly varying disturbances that do not satisfy Assumption \ref{assum3}. 
\end{remark}
The discrete-time representation of the system \eqref{sys1} using the forward shift operator $q$, which is defined as $q\xi\left(k\right) \triangleq \xi\left(k+1\right)$, at the sampling period $\tau$ is given by
\vspace{-0.33in}

\begin{align}
\begin{split}
\label{sys2}
q\xi\left(k\right)&=A_{\tau}\xi\left(k\right)+B_{\tau}\left(u\left(k\right)+d\left(k\right)\right),\\
y\left(k\right)&=C\xi\left(k\right).
\end{split}
\end{align}
\vspace{-0.32in}

Here we have utilized the fact that using Assumption \ref{assum3}, $d(t)=d(k)$ for all $k\tau \leq t < (k+1)\tau$ where $k \in \mathbb{Z}_{\geq 0}$. The system matrix and the control matrix are
\vspace{-0.25in}

\begin{equation}
\label{eq1}
A_{\tau}=\exp\left({A\tau}\right), \ \mbox{and} \
B_{\tau}=\int\limits_{0}^{\tau} \exp\left({As}\right)\, \mathrm{d}s B,
\end{equation}
\vspace{-0.25in}

respectively. By the abuse of notation, we denote $\xi(k) \coloneqq \xi(k\tau)$ for a given sampling period $\tau>0$. It is evident from \eqref{eq1} that the discrete-time representation of the system using the shift operator becomes numerically ill-conditioned when $\tau\to 0$ i.e., $\lim _{\tau \to 0}A_{\tau}=I$ and $\lim _{\tau \to 0}B_{\tau}=0$. 
\vspace{-0.05in}

\quad To circumvent the above problem, the system \eqref{sys1} is represented using the delta operator, sampled at $\tau$ interval as follows:
\vspace{-0.38in}

\begin{align}
\begin{split}
\label{sys3}
\delta \xi\left(k\right)&=A_{\delta\tau} \xi\left(k\right)+B_{\delta\tau} \left(u\left(k\right)+ d\left(k\right)\right), \\
y\left(k\right)&=C_{\delta\tau} \xi\left(k\right),
\end{split}
\end{align}
\vspace{-0.38in}

where
\vspace{-0.38in}

\begin{align}
\label{eq2}
\begin{split}
A_{\delta\tau} &= \dfrac{A_\tau-I}{\tau}=\left(I+\frac{A\tau}{2!}+\frac{A^2\tau^2}{3!}+\dotsb\right)A, \\
B_{\delta\tau} &=\dfrac{B_\tau}{\tau}=\left(I+\frac{A\tau}{2!}+\frac{A^2\tau^2}{3!}+\dotsb\right)B, \ \mbox{and} \ C_{\delta\tau} =C 
\end{split}
\end{align}
\vspace{-0.3in}

for $\tau\neq 0$. It can be seen from \eqref{eq2} that $\lim _{\tau \to 0}A_{\delta\tau}=A$ and $\lim _{\tau \to 0}B_{\delta\tau}=B$. This implies that the discrete-time representation of the system using the delta operator mimics the continuous time system when the sampling period tends to zero. Without loss of generality, we call the discrete-time representation of the system using shift operator and delta operator as the discrete-time system and the delta operator system respectively. 
\vspace{-0.05in}

\begin{figure}[h]
\centering
\includegraphics[width=\linewidth]{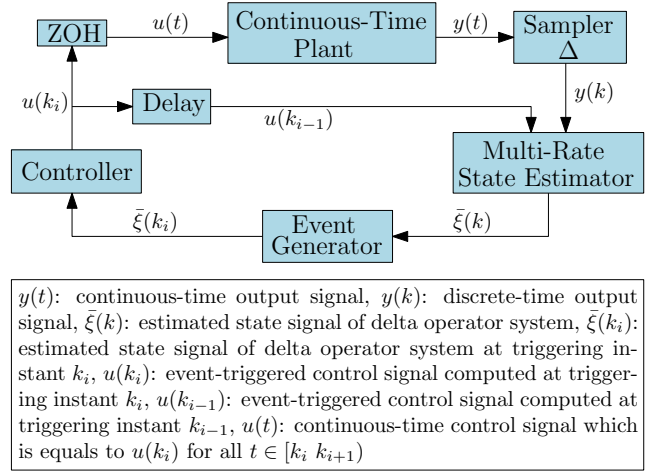}
\small\caption{System architecture}
\label{sysarc}
\end{figure}

\quad The main interest of the paper is stabilization of the delta operator system. However, by selecting an appropriate sampling period $\tau$, we can say that control designed using the delta operator system would also stabilize the continuous-time system. The schematic for the same is given in Fig. \ref{sysarc}. The delta operator system lies inside the MRSE block and the controller block computes the event-triggered control law proposed later in this paper. 
\begin{remark}
The delta operator system is said to be stable if all the eigenvalues of the system matrix lie in the circle of radius $1/\tau$ centered at $(-1/\tau,0)$ in the complex plane.
\end{remark}
\vspace{-0.05in}

\begin{lemma} \cite{uniapp}
\label{lemma1}
Given two time functions, the delta operator has the following property for a given sampling period $\tau$
\vspace{-0.08in}
\begin{align*}
\delta\left(x\left(t\right)y\left(t\right)\right) = \delta\left(x\left(t\right)\right)y\left(t\right)+x\left(t\right)\delta\left(y\left(t\right)\right)+\tau \delta\left(x\left(t\right)\right)\delta\left(y\left(t\right)\right).
\end{align*}
\end{lemma}
\vspace{-0.05in}

\subsection{Design of SMC for delta operator system}
\vspace{-0.10in}

The design of SMC involves two steps. The first step is the design of a sliding variable and the second step is to design a stabilizing controller. It is to be noted that in SMC design, the system is transformed into regular form. For this, the input matrix $B_{\delta \tau}$ can be partitioned as $B_{\delta \tau}=[B_{\delta \tau1}^{\top} \ B_{\delta \tau2}^{\top}]^{\top}$, where $B_{\delta \tau2}$ is invertible. Thus for the system like \eqref{sys3}, there exists a nonsingular transformation matrix $T \in \mathbb{R}^{n \times n} = \begin{bmatrix}
I_{(n-1)\times (n-1)} & -B_{\delta\tau1}B_{\delta \tau2}^{-1} \\
0_{1\times (n-1)}  & I_{1 \times 1}
\end{bmatrix}$ such that $x(k) = T \xi(k)$ and the system in regular form is represented as 
\vspace{-0.08in}
\begin{align}
\label{trsys1}
\begin{split}
\delta{x}(k) &= \bar{A}x(k)+\bar{B}\left(u(k)+d(k)\right), \\
y(k) &= \bar{C}x(k),
\end{split}
\end{align}
\vspace{-0.3in}

where $\bar{A} = TA_{\delta\tau} T^{-1}$, $\bar{B} = TB_{\delta\tau} $ and $\bar{C} = C_{\delta\tau}T^{-1}$. Let the sliding variable is defined as
\vspace{-0.20in}

\begin{equation}
\label{eq4}
s(k)=c^{\top}x\left(k\right),
\end{equation}
\vspace{-0.25in}

where $c=[c_1^\top \ 1]^\top \in \mathbb{R}^n$, is the design parameter called the \textit{sliding surface parameter}. Define sliding manifold as
\vspace{-0.25in}

\begin{equation*}
\mathcal{S} \triangleq \left\{x \in \mathbb{R}^n : s= c^\top x=0 \right\}.
\end{equation*}
\vspace{-0.25in}

In DTSM, control is applied to the system periodically at sampling intervals and due to this discrete nature of the control implementation, the sliding variable does not go to zero. Therefore, the trajectories of the system do not slide on the sliding manifold, and rather get attracted to the band around the sliding manifold. This band is defined as the Quasi-sliding mode band (QSMB) in literature \cite{var}.
\begin{definition} \cite{novel} (Quasi-Sliding Mode)
The system \eqref{trsys1} is said to be in \it{quasi-sliding mode} if there exists a $\bar{k} \geq 0$ such that 
\vspace{-0.15in}
\begin{align}
\label{band1}
\left| s(k)\right| \leq \Omega, \quad \forall \ k\geq \bar{k}
\end{align}
\vspace{-0.4in}

for some constant $\Omega>0$. In this case, $\Omega$ is called a \it{quasi-sliding mode band} (QSMB).
\end{definition}
It is well known in DTSM literature that a reaching law is needed to design SMC. Here we use the constant rate reaching law given for the delta operator system in \cite{dek} as
\vspace{-0.41in}

\begin{align}
\label{eq7}
\delta s(k)=-\epsilon \operatorname{sgn}s(k)+\tilde{d}(k),
\end{align}
\vspace{-0.4in}

where $\epsilon$ is the switching gain and $\tilde{d}(k)=c^\top \bar{B}d(k)$ with $\sup_{k \in \mathbb{Z}_{\geq 0}} \lvert\tilde{d}(k)\rvert\leq d_m$. Using the above reaching law, we can obtain a control input as
\vspace{-0.2in}

\begin{equation}
\label{control}
u(k)=-(c^\top \bar{B})^{-1}\left(c^\top \bar{A} x(k)+\epsilon \operatorname{sgn}s(k)\right),
\end{equation}
\vspace{-0.27in}

where $\epsilon>d_m$. This control law ensures that the sliding mode exists in the system \eqref{trsys1} and the state trajectory is bounded in a band around the sliding manifold. Readers are referred to \cite{dek} to see the proof of the above result. 
\begin{remark}
The constant rate reaching law \eqref{eq7} for the delta operator system is motivated from the reaching law approach used to design SMC in the literature and it has a fair agreement with its discrete and continuous counterparts. The term $\tilde{d}(k)$ has been added to deal with the disturbance in the system and it is a standard approach in the SMC literature \cite{jana3}.
\end{remark}
\vspace{-0.05in}

\subsection{MRSE for discrete-time system}
\vspace{-0.10in}

The concept of MROF is to sample the input and the output of the system at different sampling rates. In MRSE technique (i.e., fast output sampling, which is one type of MROF technique), the output of the system is sampled at a faster rate than the input. In one input sampling period $\tau$, the output is sampled $N$ times at sampling period $\Delta$ and these output samples are used to estimate the states at $\tau$ instants. This results in representing the state of the system in terms of the past control input and multi-rate sampled output of the system. For this, the system \eqref{sys1} is discretized at the sampling period $\Delta = \tau / N$, where $N$ is the integer greater than or equal to the rank of the observability matrix of the $\Delta$ sampled system, and it is represented as
\vspace{-0.3in}

\begin{align*}
\begin{split}
\xi(j+1)&=A_{\Delta} \xi(j)+B_{\Delta}\left(u(j)+d(j)\right), \\
y(j)&=C_{\Delta}\xi(j),
\end{split}
\end{align*}
\vspace{-0.3in}

where $A_{\Delta}=\exp({A\Delta}), \ B_{\Delta}=\int\limits_{0}^{\Delta} \exp({As})\, \mathrm{d}s B$. With a little abuse of notation, we denote $\xi(j) \coloneqq \xi(j\Delta)$ for a given $\Delta>0$. Since $\Delta=\tau/N$, if $j$ becomes a $k$ multiple of $N$ (i.e., $j = kN$), then we have $x(j)=x(kN\Delta)=x(k\tau)=x(k)$, which is consistent. Moreover, using Assumption \ref{assum3}, we can say that $d(j)=d(k)$ for all $j$ such that $k\tau \leq j\Delta < (k+1)\tau$. The fast output sampling results in following equations
\vspace{-0.3in}

\begin{align}
\nonumber \xi(k+1) &= A_{\tau}\xi(k)+B_{\tau}\left(u(k)+d(k)\right),\\
\label{sys4}
y_{k+1} &= C_{o}^{d}\xi(k)+D_{o}^{d}\left(u(k)+d(k)\right),
\end{align}
\vspace{-0.3in}

where 

\vspace{-0.4in}
\begin{small}
\begin{align*}
C_{o}^{d}= \begin{bmatrix}
C \\
CA_{\Delta} \\
\vdots \\
C A_{\Delta}^{N-1}
\end{bmatrix}, \ D_{o}^{d} = \begin{bmatrix}
0 \\
CB_{\Delta}\\
\vdots \\
C \sum_{i=0}^{N-2}A_{\Delta}^{i}B_{\Delta}
\end{bmatrix}, \ y_k :=  \begin{bmatrix}
y\left((k-1)\tau\right) \\
y\left((k-1)\tau+\Delta\right) \\
\vdots \\
y(k\tau-\Delta)
\end{bmatrix},
\end{align*}
\end{small}
\vspace{-0.4in}

$$A_{\tau}=A_{\Delta}^{N}, \quad \mbox{and} \quad B_{\tau}= \sum_{i=0}^{N-1}A_{\Delta}^{i}B_{\Delta}.$$ Here $y_k$ represents the stack of past multi-rate output samples and $C_{o}^{d}$ is the observability matrix of the $\Delta$ sampled discrete-time system. The algebraic relation among the state, past output vector, past control input, and the disturbance signal is given by
\vspace{-0.25in}

\begin{equation}
\label{eq11}
\xi(k)=L_{y}^{d}y_k+L_{u}^{d}\left(u(k-1)+d(k-1)\right),
\end{equation}
\vspace{-0.28in}

where
\vspace{-0.35in}

\begin{align*}
L_{y}^{d}=A_{\tau}\left(C_{o}^{d^\top}C_{o}^{d}\right)^{-1}C_{o}^{d^{\top}},\ L_{u}^{d}=B_{\tau}-L_{y}^{d}D_{o}^{d}.
\end{align*}
\vspace{-0.35in}

The relation obtained in \eqref{eq11} has the disturbance term $d(k-1)$ and the exact information of disturbance is unknown. So it is not possible to design the control using this estimated state. Thus we represent the state estimate as
\vspace{-0.35in}

\begin{align}
\label{eseq}
\bar{\xi}(k)=L_{y}^{d}y_k+L_{u}^{d}u(k-1).
\end{align}
\vspace{-0.35in}

The above state estimation technique may not work properly with a very small sampling period in the numerically ill-conditioned case of the matrix $(C_{o}^{d^\top}C_{o}^{d})^{-1}$, which would limit the application of MRSE based approach. Therefore, in the next section, a new observability matrix for MRSE is designed to solve the above problem.
\vspace{-0.05in}

\section{MRSE based SMC for delta operator system}  \label{mr1}
\vspace{-0.10in}

This section presents one of the main results of the paper. It includes the proposal of a new observability matrix and an MRSE technique for the delta operator system. This is followed by the design of SMC based on the proposed MRSE technique.
\vspace{-0.05in}

\subsection{MRSE for delta operator system}
\vspace{-0.10in}

The formulation of the MRSE technique for the delta operator system requires the system to be sampled at two different sampling rates. For the system \eqref{sys1}, the $\tau$ sampled delta operator system is given in \eqref{sys3} and the $\Delta$ sampled delta operator system would be
\vspace{-0.06in}
\begin{align*}
\begin{split}
\delta \xi(j)&=A_{\delta\Delta} \xi(j)+B_{\delta\Delta} \left(u(j)+ d(j)\right), \\
y(j)&=C_{\delta\Delta} \xi(j),
\end{split}
\end{align*}
\vspace{-0.27in}

where $A_{\delta\Delta} = (A_\Delta-I)/ \Delta$, $B_{\delta\Delta}=B_\Delta/ \Delta$ and $C_{\delta\Delta} =C$. To tackle the numerical singularity problem of discrete-time observability matrix $C_{o}^{d}$, define a new observability matrix as
\vspace{-0.22in}

\begin{equation}
\label{limeq}
C_{o}^{\delta}=\begin{bmatrix}
C \\
CA_{\delta\Delta} \\
\vdots \\
CA_{\delta\Delta}^{N-1}
\end{bmatrix}.
\end{equation}
As $\Delta$ tends to $0$, the matrix $C_{o}^{\delta}$ converges to continuous-time observability matrix $C_o$ i.e., 
\vspace{-0.10in}
\begin{equation}
\label{limeq1}
\lim _{\Delta \to 0}C_{o}^{\delta}= C_o= \begin{bmatrix}
C \\ CA \\\vdots \\ CA^{N-1}
\end{bmatrix}.
\end{equation}
The explicit relation between $C_{o}^{d}$ and $C_{o}^{\delta}$ is given in the following theorem.
\begin{theorem} \label{th1}
Given the matrices $C_{o}^{d}$ and $C_{o}^{\delta}$, it holds that, for any $\Delta >0$,
\vspace{-0.05in}
\begin{align}
\label{rel}
C_{o}^{\delta}=E_pQ_pC_{o}^{d}
\end{align}
\vspace{-0.3in}

where
$E_p = \mbox{blockdiag} \left\{ I_p, \frac{I_p}{\Delta}, \cdots, \frac{I_p}{\Delta^{N-1}} \right\}$ with the $p\times p$ identity matrix $I_p$, and
\vspace{-0.3in}

\begin{small}
\begin{align*}
Q_p = \begin{bmatrix}
I_p \\
-I_p &  I_p  \\
I_p &  -2I_p & I_p \\
-I_p & 3I_p & -3I_p & I_p \\
\vdots & \vdots & \vdots & \vdots &\ddots \\
r_{N,1}I_p & r_{N,2}I_p & \cdot & \cdot & \cdots & I_p
\end{bmatrix}
\end{align*}
\end{small}
\vspace{-0.20in}

in which, the coefficient $r_{i,j}$ in the lower (off-) diagonal position is given by, for $1\leq l \leq N$ and $1 \leq k \leq l$,
\vspace{-0.32in}

\begin{align*}
r_{l,k}=\dfrac{1}{(k-1)!}\dfrac{\mathrm{d}^{k-1}}{\mathrm{d}s^{k-1}}f_{l-1}(s)\Big|_{s=0},
\end{align*}
\vspace{-0.25in}

where $f_{l-1}(s)=\left(s-1\right)^{l-1}$ and the upper diagonal terms are all zeros.
\end{theorem}
\vspace{-0.15in}

\begin{pf} \label{pf1}
We can prove this theorem by using the fact that
\vspace{-0.35in}

\begin{align}
\label{fac}
\begin{split}
I_n &= I_n\\
A_{\Delta}-I_n &= -I_n+A_{\Delta}\\
\left(A_{\Delta}-I_n\right)^2 &= I_n-2A_{\Delta}+A_{\Delta}^2\\
\left(A_{\Delta}-I_n\right)^3 &= -I_n+3A_{\Delta}-3A_{\Delta}^{2}+A_{\Delta}^3\\
\vdots \qquad &= \qquad \vdots
\end{split}.
\end{align}
\vspace{-0.3in}

Now using $Q_p$, $E_p$ and \eqref{fac} in relation \eqref{rel}, the right-hand side of \eqref{rel} results in \eqref{limeq}. It is noted that, the matrix $C_{o}^{\delta}$ becomes almost invariant to sufficiently small $\Delta$ because of \eqref{limeq}. Observing that rank$\left(C_{o}^{\delta}\right)$=rank$\left(C_{o}^{d}\right)$ since $E_p$ and $Q_p$ are invertible, the discrete-time observability check can be robustly performed with $C_{o}^{\delta}$ instead of $C_{o}^{d}$, in particular, when the sampling period is small. \hfill \qed
\end{pf}
\begin{remark}
In a similar context, the conventional discrete-time controllability matrix
\vspace{-0.05in}
\begin{equation*}
\mathcal{C}_r^{d} =\left[B_{\Delta} \quad A_{\Delta}B_{\Delta} \quad \cdots \quad A_{\Delta}^{N-1}B_{\Delta}\right]
\end{equation*}
\vspace{-0.32in}

would be numerically rank-deficient as $\Delta \rightarrow 0.$ Suppose that the controllability matrix $\mathcal{C}_r^{\delta}$ is defined by replacing $\left(A_{\Delta},B_{\Delta}\right)$ with $\left(A_{\delta \Delta},B_{\delta \Delta}\right)$ in $\mathcal{C}_r^{d}$. Then, by utilizing the relationship in \eqref{rel}, it can be shown that
\vspace{-0.4in}

\begin{align*}
\frac{1}{\Delta}\mathcal{C}_r^{d}.Q_{m}^{T}.E_{m}^{T} = \mathcal{C}_r^{\delta},
\end{align*}
\vspace{-0.3in}

which leads to $rank\left(\mathcal{C}_r^{d}\right)=rank\left(\mathcal{C}_r^{\delta}\right)$. Considering that $\mathcal{C}_r^{\delta}$ approaches the continuous-time controllability matrix as $\Delta \rightarrow 0$, $\mathcal{C}_r^{d}$ is a more robust index for the rank test.
\end{remark}
Now, we propose an improved scheme for MRSE by making use of the key relationship in \eqref{rel}. Multiplying relation \eqref{sys4} by the matrix, $E_pQ_p$, we get
\vspace{-0.35in}

\begin{align*}
E_pQ_py_{k+1} &= E_pQ_pC_{o}^{d}\xi(k)+E_pQ_pD_{o}^{d}\left(u(k)+d(k)\right),\\
&= C_{o}^{\delta}\xi(k)+D_{o}^{\delta}\left(u(k)+d(k)\right),
\end{align*}
where
\begin{equation*}
D_{o}^{\delta}:=E_pQ_pD_{o}^{d}=\begin{bmatrix}
0 \\
CB_{\delta\Delta}  \\
\vdots \\
C\left( A_{\delta\Delta} \right)^{N-2}B_{\delta\Delta}
\end{bmatrix}.
\end{equation*}
\vspace{-0.150in}

Thus we have the estimated state of the delta operator system using MRSE as 
\vspace{-0.25in}

\begin{equation}
\label{estate}
\xi(k)=L_{y}^{\delta}y_{k}+L_{u}^{\delta}\left(u(k-1)+d(k-1)\right),
\end{equation}
where 
\vspace{-0.1in}
\begin{align*}
L_{y}^{\delta}&=A_{\tau}\left(C_{o}^{\delta^T}C_{o}^{\delta}\right)^{-1}C_{o}^{\delta^T}E_pQ_p, \\
L_{u}^{\delta}&=B_{\tau}-A_{\tau}\left(C_{o}^{\delta^T}C_{o}^{\delta}\right)^{-1}C_{o}^{\delta^T}D_{o}^{\delta}.
\end{align*}
\vspace{-0.35in}

Comparing with the expression in \eqref{eq11}, one may observe that the numerical singularity in the matrix inversion can be avoided by using $C_{o}^{\delta}$ for fast sampling cases, unless the continuous-time observability matrix is ill-conditioned. The estimated state in relation \eqref{estate} needs information of disturbance so we define a new relationship after discarding uncertainty as
\vspace{-0.22in}

\begin{equation}
\label{est2}
\bar{\xi}(k)=L_{y}^{\delta}y_{k}+L_{u}^{\delta}u(k-1).
\end{equation}
\vspace{-0.25in}

The design of control \eqref{control} needs estimated states, so define 
\vspace{-0.35in}

\begin{align}
\label{estst1}
\bar{x}(k) := T \bar{\xi}(k) = TL_{y}^{\delta}y_{k} + TL_{u}^{\delta}u(k-1).
\end{align} 
\vspace{-0.3in}

As a result, we obtain the relation $x(k) = \bar{x}(k)+ TL_{u}^{\delta}d(k-1)$. The relation \eqref{estst1} is independent of disturbance, and it is used further to design the event-triggered control law. But as the state is approximated to discard the effect of disturbance and not exactly known, so there is also an uncertainty in the measurement of the sliding variable. Therefore, we define
\vspace{-0.22in}

\begin{equation}
\label{sbar}
s(k) := \bar{s}(k) +l(k-1),
\end{equation}
\vspace{-0.25in}

where $\bar{s}(k)= c^{\top} \bar{x}(k) = c^{\top}TL_{y}^{\delta}y_{k} + c^{\top} TL_{u}^{\delta}u(k-1)$ is sliding variable designed using estimated state $\bar{x}(k)$ and $l(k)= c^{\top}TL_{u}^{\delta}d(k)$ with $ \sup_{k \in \mathbb{Z}_{\geq 0}} \lvert l(k) \rvert  \leq l_m$. Now to design the control law, only the sign of $s(k)$ is needed and it can be noted from \eqref{sbar} that $\operatorname{sgn}s(k)=\operatorname{sgn}\bar{s}(k)$ if $|s(k)|>l_m$. Hence the sign of $s(k)$ is correctly obtained when $\left|s(k)\right|>l_m$, so $\operatorname{sgn}s(k)$ is replaced with $\operatorname{sgn}\bar{s}(k)$ in control input. For the case $|s(k)|<l_m$, the sign of $s(k)$ cannot be determined accurately, so its effect is taken care in the stability analysis. The design of SMC using the estimated state is given in following.
\vspace{-0.05in}

\subsection{Design of SMC using MRSE technique}
\vspace{-0.10in}

To design SMC using the estimated state we modify the reaching law by replacing $s(k)$ with $\bar{s}(k)$ in the signum function as
\vspace{-0.38in}

\begin{align}
\label{eq6}
\delta s(k)=-\epsilon \operatorname{sgn}\bar{s}(k)+\tilde{d}(k)+f(k-1),
\end{align}
\vspace{-0.38in}

where $f(k)=c^\top \bar{A}TL_{u}^{\delta}d(k)$ with $\sup_{k \in \mathbb{Z}_{\geq 0}} \lvert\ f(k)\rvert\leq f_m$. Applying the delta operator on sliding variable gives $\delta s(k)= \left(s(k+1)-s(k)\right)/ \tau $ and using equations \eqref{trsys1} and \eqref{eq4} in this relation yields
\vspace{-0.35in}

\begin{align}
\label{eq8} \delta s(k)&=c^\top \bar{A}x(k)+c^\top \bar{B}u(k)+c^\top \bar{B}d(k).
\end{align}
\vspace{-0.35in}

Using $x(k)=\bar{x}(k)+TL_{u}^{\delta}d(k-1)$, and the reaching law \eqref{eq6} we obtain the control law as
\vspace{-0.20in}

\begin{equation}
\label{control1}
u(k)=-(c^\top \bar{B})^{-1}(c^{\top}\bar{A}\bar{x}(k)+\epsilon \operatorname{sgn}\bar{s}(k) )
\end{equation}
\vspace{-0.25in}

which ensures that the sliding mode exists in the system. We prove the same in the theorem below.
\begin{theorem} \label{th2}
Consider the system \eqref{trsys1}, sliding variable \eqref{eq4} and reaching law \eqref{eq6}. Let the control input \eqref{control1} be applied to the system. Then the quasi-sliding mode occurs in the system if the switching gain is selected as $\epsilon>d_m+f_m$.
\end{theorem}
The proof of above theorem is given in Appendix \ref{app1}. The system \eqref{trsys1} can be rewritten in the regular form as
\vspace{-0.32in}

\begin{subequations}
\begin{align}
\label{regform1}
\delta{x}_1(k) &= \bar{A}_{11}x_1(k)+\bar{A}_{12}x_2(k), \\
\label{regform2}
\delta{x}_2(k) &= \bar{A}_{21}x_1(k)+\bar{A}_{22}x_2(k)+\bar{B}_{2}\left(u(k)+d(k)\right).
\end{align}
\end{subequations}
\vspace{-0.32in}

From \eqref{eq4}, we can write $ x_2(k)=-c_1^\top x_1(k)+s(k).$ Substituting this into \eqref{regform1} results in \vspace{-0.32in}

\begin{align*}
\delta x_1(k)=(\bar{A}_{11}-\bar{A}_{12}c_1^\top)x_1(k)+\bar{A}_{12}s(k).
\end{align*}
\vspace{-0.32in}

Choosing $c_1$ such that the eigenvalues of the matrix $A_{cl}=\bar{A}_{11}-\bar{A}_{12}c_1^\top$ are placed in the circle of radius $1/\tau$ centered at $(-1/\tau,0)$ in the complex plane guarantees boundedness of $x_1(k)$. The calculation of the bound of state trajectories is given in the following proposition.
\begin{proposition} \label{pro1}
Consider the system \eqref{trsys1} and QSMB \eqref{band2}. The system trajectories remain bounded in the region given by 
\vspace{-0.4in}

\begin{align}
\nonumber \Theta_1 = & \Bigg\lbrace   x \in \mathbb{R}^{n}:\| x \|\leq (1+\| c_1 \|) \\ \label{band5} & \times \left(\dfrac{\lambda_{\max}(P)(\sqrt{c_2}+b_2)^2+\tau a_2c_2}{\lambda_{\min}(P)}\right)^{1/2}+\Omega \Bigg\rbrace
\end{align}
\vspace{-0.3in}

where $\gamma_1 = \left\|P \bar{A}_{12}+\tau A_{cl}^{\top}P \bar{A}_{12} \right\| \Omega$, $\gamma_2 = \tau \lambda_{\max}(P) \left\| \bar{A}_{12}\right\|^2 \Omega^2$, $a_2 = \lambda_{\min}(Q) $, $b_2 =\gamma_1/a_2 $, $c_2 = \gamma_2/a_2+b_2^2 $, $P$ and $Q$ are positive definite and satisfy $A_{cl}^{\top}P+PA_{cl}+\tau A_{cl}^{\top}PA_{cl}=-Q$.
\end{proposition}
\vspace{-0.15in}

\begin{pf}  
Considering the Lyapunov function $V_1(x_1(k))=x_1^{\top}(k)Px_1(k)$, and using Lemma \ref{lemma1} we obtain
\vspace{-0.33in}

\begin{align*}
\delta V_1(x_1(k)) &= \delta x_1^{\top}(k)Px_1(k)+x_1^{\top}(k)P\delta x_1(k)+\tau \delta x_1^{\top}(k)P\delta x_1(k) \\
              &= x_1^{\top}(k)\left(A_{cl}^{\top}P+PA_{cl}+\tau A_{cl}^{\top}PA_{cl}\right)x_1(k) \\
               &\ +2x_1^{\top}\left(k\right)\left(P \bar{A}_{12}+\tau A_{cl}^{\top}P \bar{A}_{12}\right)s(k)+\tau s^2(k)\bar{A}_{12}^{\top}P\bar{A}_{12}.
\end{align*}
\vspace{-0.34in}

After reaching QSMB, we have $|s(k)| < \Omega$, using this in the above equation gives
\vspace{-0.33in}

\begin{align*}
\delta V_1(x_1(k)) \leq & -\lambda_{\min}(Q)\left\|x_1(k) \right\|^2 +\tau \lambda_{\max}(P) \left\| \bar{A}_{12}\right\|^2 \Omega^2 \\
               &+2\Omega\left\|P \bar{A}_{12}+\tau A_{cl}^{\top}P \bar{A}_{12} \right\| \left\| x_1(k)\right\| \\
               =& -a_2 \left(\left(\left\| x_1(k)\right\|-\dfrac{\gamma_1}{\lambda_{\min}(Q)}\right)^2   - \dfrac{\gamma_1^2+\lambda_{\min}(Q)\gamma_2}{\lambda_{\min}^2(Q)}\right) \\
               =&-a_2\left(\left(\left\|x_1(k)\right\| - b_2\right)^2-c_2 \right).
\end{align*}
\vspace{-0.33in}

From the last inequality, it can be seen that whenever $\|x_1(k) \| > \sqrt{c_2}+b_2$,  $\delta V_1(k) < 0$. So the maximum deviation in state variable in one time step is calculated using last inequality as  $\left(\left(\lambda_{\max}(P)(\sqrt{c_2}+b_2)^2+\tau a_2c_2\right)/\left(\lambda_{\min}(P)\right)\right)^{1/2}$. Hence $x_1(k)$ becomes bounded in the region \\
$\left\lbrace x_1 \in \mathbb{R}^{n-1}: \| x_1\| \leq \left(\frac{\lambda_{\max}(P)(\sqrt{c_2}+b_2)^2+\tau a_2c_2}{\lambda_{\min}(P)}\right)^{1/2}\right\rbrace$ for all $k>\bar{k}$. It further results in
\begin{align*}
\| x(k)\| & \leq \left\|x_1(k)\right\| +\left| x_2(k) \right|  \leq \left(1+\|c_1 \|\right)\| x_1(k) \| +\Omega.
\end{align*}
\vspace{-0.33in}

The last relation is obtained using \eqref{eq4} and \eqref{band1}. Therefore, the set $\Theta_1$ in which system trajectories remain bounded is given by \eqref{band5}. This completes the proof. \hfill \qed
\end{pf}
\vspace{-0.15in}

Since the control is designed using states estimated through MRSE technique, the calculation of control input at the initial time $t=0$ needs $\bar{x}(0)$ and this estimation requires output information prior to time $t=0$. But the system starts to evolve only after $t=0$. Hence the calculation of initial control is not possible. Therefore, there is an error in the control signal at $t=0$. We can represent the control signal \eqref{control1} as
\vspace{-0.33in}

\begin{align}
\label{con1}
u(k) = F\bar{x}(k)+g_1(k),
\end{align}
\vspace{-0.33in}

where $F=-(c^\top \bar{B})^{-1}c^{\top}\bar{A}$ and $g_1(k) = -(c^\top \bar{B})^{-1} \epsilon \operatorname{sgn} \bar{s}(k).$ Now assuming some $\hat{x}(0)$, the control signal at the initial time is calculated as $u(0)=F\hat{x}(0)+g_1(0)$ which is not the actual control. So we define the error in the control signal as
\vspace{-0.21in}

\begin{equation}
\label{conerr}
e_u(k)=u(k)-F\bar{x}(k)-g_1(k).
\end{equation}
\vspace{-0.24in}

The stability of the error dynamics resulting from \eqref{conerr} is discussed in \cite{hwerner}. But the control law proposed in this paper differs from the one in \cite{hwerner}. Thus to prove the stability of the error dynamics, we adopt a similar approach to \cite{hwerner} which is given by the following lemma. Note that $e_u(0)\neq 0$ but $e_u(k)=0$ for all $k \in \{ \mathbb{Z}_{\geq 0} \setminus 0\} $ and same is shown below.
\begin{lemma} \label{le1}
Consider the system \eqref{trsys1}, control input \eqref{control1} and estimated state \eqref{estst1}. Given any initial condition $\bar{x}(0)$, the error introduced in the control signal at the initial time becomes zero after one-time step.
\end{lemma}
\vspace{-0.05in}

The proof of this Lemma is given in Appendix \ref{app3}.
\vspace{-0.05in}

\section{Event-triggered SMC based on new MRSE technique}  \label{mr2}
\vspace{-0.10in}

In this section, to achieve minimum resource utilization with robustness, MRSE based event-triggered SMC is presented for the delta operator system. Later a triggering rule is also provided for generating the triggering instants.  
\vspace{-0.05in}

\subsection{Design of MRSE based event-triggered SMC}
\vspace{-0.10in}

Here to obtain the minimum number of control updates along with the advantage of MRSE technique, the design of event-triggered SMC for delta operator system using proposed MRSE technique is proposed. Ideally in a continuous-time system, the state slides on the sliding manifold, hence the band size is zero. But the application of control in an event-triggering manner yields the practical sliding mode. This differs from the discrete-time system case, where an inherent band called QSMB already exists because of the discrete nature of the controller. However, application of event-triggered strategy to the discrete representation of the system results in the practical quasi-sliding mode.  
\begin{definition} \cite{event2} (Practical Quasi-Sliding Mode)
The system \eqref{trsys1} is said to be in practical quasi-sliding mode if given some $\Omega_1 >0$, there exists a $\hat{k} \geq 0$, such that 
\vspace{-0.34in}

\begin{align}
\label{band3}
\left| s(k)\right| \leq \Omega_1 \quad \forall \ k\geq \hat{k}.
\end{align}
\vspace{-0.34in}

The constant $\Omega_1$ is called a practical QSMB in the vicinity of the sliding manifold.
\end{definition}
\begin{remark}
The size of QSMB depends on sampling period whereas the size of practical QSMB depends on both the triggering parameter and sampling period. So for a fixed sampling period the size of practical QSMB can be varied using the triggering parameter. The size of practical QSMB is greater than or equals to the size of QSMB and this is proved later in the paper.
\end{remark}
Let there exists a sequence of triggering instants which is denoted by $\left\{ k_i\right\}_{i\in \mathbb{Z}_{\geq 0}}$, where $\left\{ k_0,k_1,k_2,\dotsc \right\}$ are aperiodic. If the control \eqref{control1} is updated at every $\left\{ k_i\right\}_{i\in \mathbb{Z}_{\geq 0}}$, an error is induced in the system which is defined as $\bar{e}(k)=\bar{x}(k_i)-\bar{x}(k)$, for all $k \in [k_i, k_{i+1}).$ The event-triggered SMC law is proposed as 
\vspace{-0.34in}

\begin{align}
\label{contrig}
u(k)=-(c^\top \bar{B})^{-1}\left(c^\top \bar{A} \bar{x}(k_i)+\epsilon\operatorname{sgn}\bar{s}(k_i)\right)
\end{align}
\vspace{-0.34in}

for all $k\in [k_i, k_{i+1})$. It is to be noted that the next control updating instant $k_{i+1}$ is not necessarily equal to $k+1$. The state $\bar{x}(k_i)$ used in \eqref{contrig} is estimated using relation \eqref{estst1} after replacing $u(k-1)$ by input computed at previous triggering instant $k_{i-1}$ i.e., $u(k_{i-1})$. Thus the communication is not periodic. The existence of the sliding mode in the system under the application of event-triggered control is presented in the following theorem.
\begin{theorem} \label{th3}
Consider the system \eqref{trsys1}, sliding variable \eqref{eq4} and reaching law \eqref{eq6}. Assume, for a given $\alpha >0$,
\vspace{-0.22in}

\begin{equation}
\label{eq10}
\left\|c\right\| \|\bar{A}\| \| \bar{e}(k)\| <\alpha
\end{equation}
\vspace{-0.25in}

holds for all $k\in \mathbb{Z}_{\geq 0}$ then, there exists a sequence of triggering instants $\left\{ k_i\right\}_{i\in \mathbb{Z}_{\geq 0}}$ for control input \eqref{contrig} such that practical quasi-sliding mode occurs in the closed-loop system if the switching gain satisfies $\epsilon>d_m+f_m+\alpha$.
\end{theorem}
The proof of above theorem is given in Appendix \ref{app2}. Further, the boundedness of state trajectories under the application of event-triggered SMC is proved in the following proposition.
\begin{proposition} \label{pro2}
Consider the system \eqref{trsys1} and practical QSMB \eqref{band4}. The system trajectories remain bounded in the region given by 
\vspace{-0.4in}

\begin{align}
\nonumber \Theta_2 = & \Bigg\lbrace   x \in \mathbb{R}^{n}: \| x \|\leq (1+\| c_1 \|)  \\  &\times \left(\dfrac{\lambda_{\max}(P)(\sqrt{c_4}+b_4)^2+\tau a_2c_4}{\lambda_{\min}(P)}\right)^{1/2}+\Omega_1 \Bigg\rbrace
\end{align}
\vspace{-0.30in}

where $\beta_1 = \|P \bar{A}_{12}+\tau A_{cl}^{\top}P \bar{A}_{12} \| \Omega_1$, $\beta_2 = \tau \lambda_{\max}(P) \| \bar{A}_{12}\|^2 \Omega_1^2$, $b_4 =\beta_1/a_2 $, and $c_4 = \beta_2/a_2+\beta_1^2/a_2^2 $ .
\end{proposition}
\vspace{-0.20in}

\begin{pf}
The proof is similar to the proof of Proposition \ref{pro1}. \hfill \qed
\end{pf}
\vspace{-0.05in}

\subsection{Event-triggering rule}
\vspace{-0.10in}

The stability of the system \eqref{trsys1} under the application of the event-triggered SMC \eqref{contrig} is achieved if the condition \eqref{eq10} is satisfied for all $k \geq 0.$ This event condition generates a sequence of control updating instants such that the system states do not blow in any two consecutive triggering instants. Moreover, a more potent condition than \eqref{eq10} to obtain event-triggering can be found such that the condition $ \|c\| \|\bar{A}\| \| \bar{e}(k)\| < \sigma\alpha$ is satisfied for some $\sigma \in (0, ~ 1)$. Therefore, the triggering rule is established as
\vspace{-0.25in}

\begin{equation}
\label{trig2}
k_{i+1}=\inf \lbrace k>k_i : \| c\| \| \bar{A}\|  \|\bar{e}(k)\| \geq \sigma \alpha \rbrace.
\end{equation}
\vspace{-0.25in}

A sequence of triggering instants $\lbrace k_i\rbrace_{i=1}^{\infty}$ is generated whenever the above triggering rule violates and this rule also takes care of the condition \eqref{eq10}. Moreover, control is discrete in nature so the Zeno execution of control is avoided.
\begin{remark}
Unlike periodic implementation, in the case of event-triggering, the control input is updated at aperiodic time instants. Hence the chances of growth in state trajectories of the system for one triggering interval is more in the event-triggered strategy. This means that the size of the band around the sliding manifold is more for event-triggered strategy as compared to the time triggered strategy. This is also clear from relation \eqref{band2} and \eqref{band4} that the size of the practical QSMB is always greater than or equals to that of the size of the QSMB i.e., $\Omega_1 \geq \Omega$.
\end{remark}
\vspace{-0.05in}

\begin{remark}
In event-triggering strategy, the triggering block is located at the sensor end, and the triggering rule decides the transmitting instant of state to the controller so that the control signal can be updated. Thus in this paper, the state is continuously monitored to evaluate \eqref{eq10} (or \eqref{trig2}) to realize this triggering rule. Once this condition is violated, the triggering instant is generated and the control task is executed. It may be noted that in real practice, the sensor measurements are assumed to be continuous for sufficiently small sampling intervals. So the continuous evaluation of triggering rule \eqref{eq10} is possible for practical realization.
\end{remark}
\vspace{-0.05in}

\section{Simulation results} \label{ne}
\vspace{-0.10in}

In this section, we present the simulation results of the paper. We have simulated two examples. The first example is the ball and beam system which uses Assumption \ref{assum3} for the disturbance. The second example is a numerical example taken from \cite{shimd} which does not use Assumption \ref{assum3} for the disturbance. In the second example, an equivalent matched and bounded disturbance is constructed using the method presented in \cite{shimd}.
\vspace{-0.05in}

\subsection{Example 1}
\vspace{-0.10in}

Consider the ball and beam system whose dynamics is given by \eqref{sys1} with matrices
\vspace{-0.08in}
\begin{align*}
A = \begin{bmatrix}
0 & 1 & 0 & 0\\
0 & 0 & 7 & 0\\
0 & 0 & 0 & 1\\
0 & 0 & 0 & 0
\end{bmatrix}, \ B = \begin{bmatrix}
0 \\ 0 \\ 0 \\
1
\end{bmatrix} \ \mbox{and} \ C= \begin{bmatrix}
1 & 0 & 0 & 0
\end{bmatrix}.
\end{align*}
\vspace{-0.32in}

The state $\xi = [r \ \dot{r} \ \phi \ \dot{\phi} ]^{\top}$ where $r$ is the ball position, $\dot{r}$ is ball velocity and $\phi$ is beam angle coordinate. The disturbance is taken as $0.05 \sin (t)$. The aim is to drive all states of the system to zero from a given initial condition. For MRSE technique, we assume $\tau = 10^{-4}$ and $N=4$. Hence the continuous time system is sampled at $\tau = 10^{-4}$sec and $\Delta = 2.5\times 10^{-5}$sec. The disturbance is assumed to be slowly varying during the sampling interval. 
\vspace{-0.08in}

\quad The discrete-time system and delta operator system for the sampling period $\tau$ are 
\vspace{-0.35in}

\begin{align}
\nonumber \xi(k+1) &= A_{\tau}\xi(k)+B_{\tau}(u(k)+d(k)), \\ 
\label{delexam}\mbox{and} \quad \delta\xi(k) &= A_{\delta \tau}\xi(k)+B_{\delta \tau}(u(k)+d(k)),
\end{align}
\vspace{-0.35in}

respectively. Under Assumption \ref{assum3}, we have sampled disturbance $d(k) = 0.05  \sin (k)$. The matrices $A_{\tau}$, $B_{\tau}$, $A_{\delta \tau}$ and $B_{\delta \tau} $ are calculated using expressions given in Section \ref{pre} and are given below  
\vspace{-0.35in}

\begin{align*}
\nonumber &A_{\tau} = \begin{bmatrix}
1 & \mathrm{E}{-4} &   3.5\mathrm{E}{-8} & 1.1667\mathrm{E}{-12} \\
0 & 1 & 7\mathrm{E}{-4} &  3.5\mathrm{E}{-8}\\
0 & 0 &  1 &  \mathrm{E}{-4} \\
0 & 0 &   0  & 1
\end{bmatrix}, \ B_{\tau} = \begin{bmatrix}
2.9167\mathrm{E}{-17} \\ 1.1667\mathrm{E}{-12} \\ 5\mathrm{E}{-9} \\ \mathrm{E}{-4}
\end{bmatrix},  \\ 
  &A_{\delta \tau} = \begin{bmatrix}
0 & 1 & 3.5\mathrm{E}{-4} & 1.1667\mathrm{E}{-8} \\
0 & 0 & 7      & 3.5\mathrm{E}{-4} \\
0 & 0 & 0      &   1 \\
0 & 0 & 0      &  0
\end{bmatrix}, \ B_{\delta \tau} = \begin{bmatrix}
2.9167\mathrm{E}{-13} \\ 1.1667\mathrm{E}{-8} \\ 5\mathrm{E}{-5} \\  1
\end{bmatrix}.
\end{align*}
\vspace{-0.28in}

It can be observed from the above representation that, the control matrix of the discrete-time system using the shift operator is nearly zero. Thus there is no control over the system, whereas the delta operator system imitates the continuous-time system.
\begin{table}[h]
\caption{Dependency of observability matrices on sampling periods}
\renewcommand{\arraystretch}{1.1}
\begin{tabular}{ p{1cm}|p{1cm}|p{1cm}|p{1cm}|p{1cm}|p{1cm}}
 \hline
 \hline
 $\Delta$ & - & \multicolumn{2}{c|}{0.1msec} & \multicolumn{2}{c}{0.025msec} \\
 \hline 
 $C_{o}^{X}$ & $C_{o}$ &  $C_{o}^{d}$ &  $C_{o}^{\delta}$ &  $C_{o}^{d}$ &  $C_{o}^{\delta}$ \\
 \hline
 $\eta$  &   7.0000  &  2.0000 & 7.0004 & 2.0000 & 7.0001\\
           &   7.0000  &  0.0002 & 6.9997 & 0.0001 & 6.9999\\
           &   1.0000  &  0.0000 & 1.0000 & 0.0000 & 1.0000\\
           &   1.0000  &  0.0000 & 1.0000 & 0.0000 & 1.0000\\     
 \hline
 rank & 4 & 4 & 4 & 4 & 4  \\
 \hline \hline
\end{tabular}
\label{tab}
$\eta$ is the singular value and $C_{o}^{X}$ denotes the different observability matrix.
\end{table}

\quad The singular values and ranks of the observability matrices are calculated by MATLAB\reg to see the dependencies on sampling periods, and it is shown in Table \ref{tab}. It can be observed that $C_{o}^{d}$ becomes ill-conditioned as $\Delta$ gets smaller while $C_{o}^{\delta}$ approaches $C_{o}$ that is well posed. Moreover, the MATLAB\reg function$-\mbox{rank}()-$ returns the incorrect outcome for $C_{o}^{d}$ in the case of $0.1\mbox{msec}$ and $0.025\mbox{msec}$, which shows that the conventional discrete-time observability check can fail for short sampling periods. However, $C_{o}^{\delta}$ produces the correct result as expected.
\vspace{-0.08in}

\quad The initial condition is taken as $\xi_0=[3 \ 2 \ 1 \ -1]^\top$ and the sliding parameter is chosen to be $c = [0.0114 \ 0.0943 \ 1.4999 \ 1]^\top$ so that the stability of the closed-loop system can be assured. The event parameters are $\alpha = 0.2$, $\sigma = 0.9$ and the switching control gain is $\epsilon=0.261$.
\begin{figure}[h]
\centering{\includegraphics[width=1\linewidth]{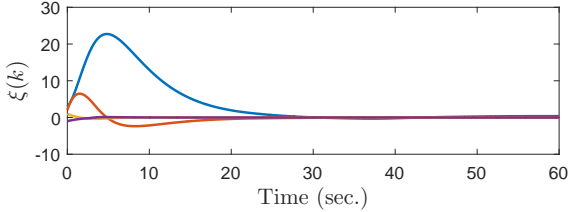}}
\small\caption{State trajectories}
\label{fig1}
\end{figure}
\begin{figure}[h]
\centering{\includegraphics[width=1\linewidth]{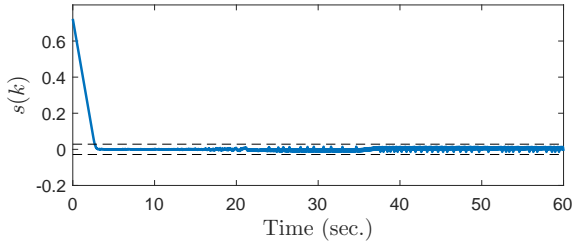}}
\small\caption{Sliding trajectory}
\label{fig2}
\end{figure}

\begin{figure}[h]
\centering{\includegraphics[width=1\linewidth]{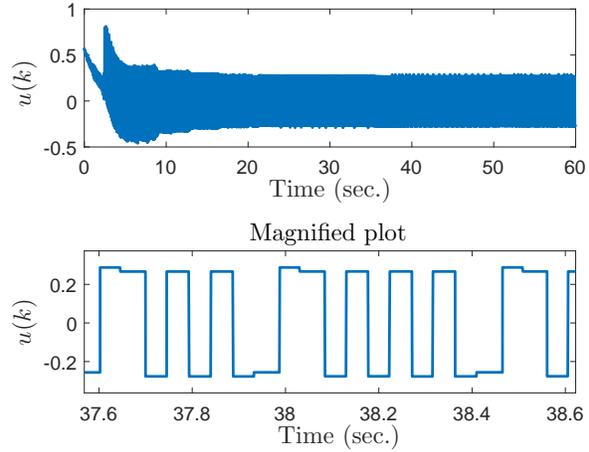}}
\small\caption{Control input}
\label{fig3}
\end{figure}

\begin{figure}[h]
\centering{\includegraphics[width=1\linewidth]{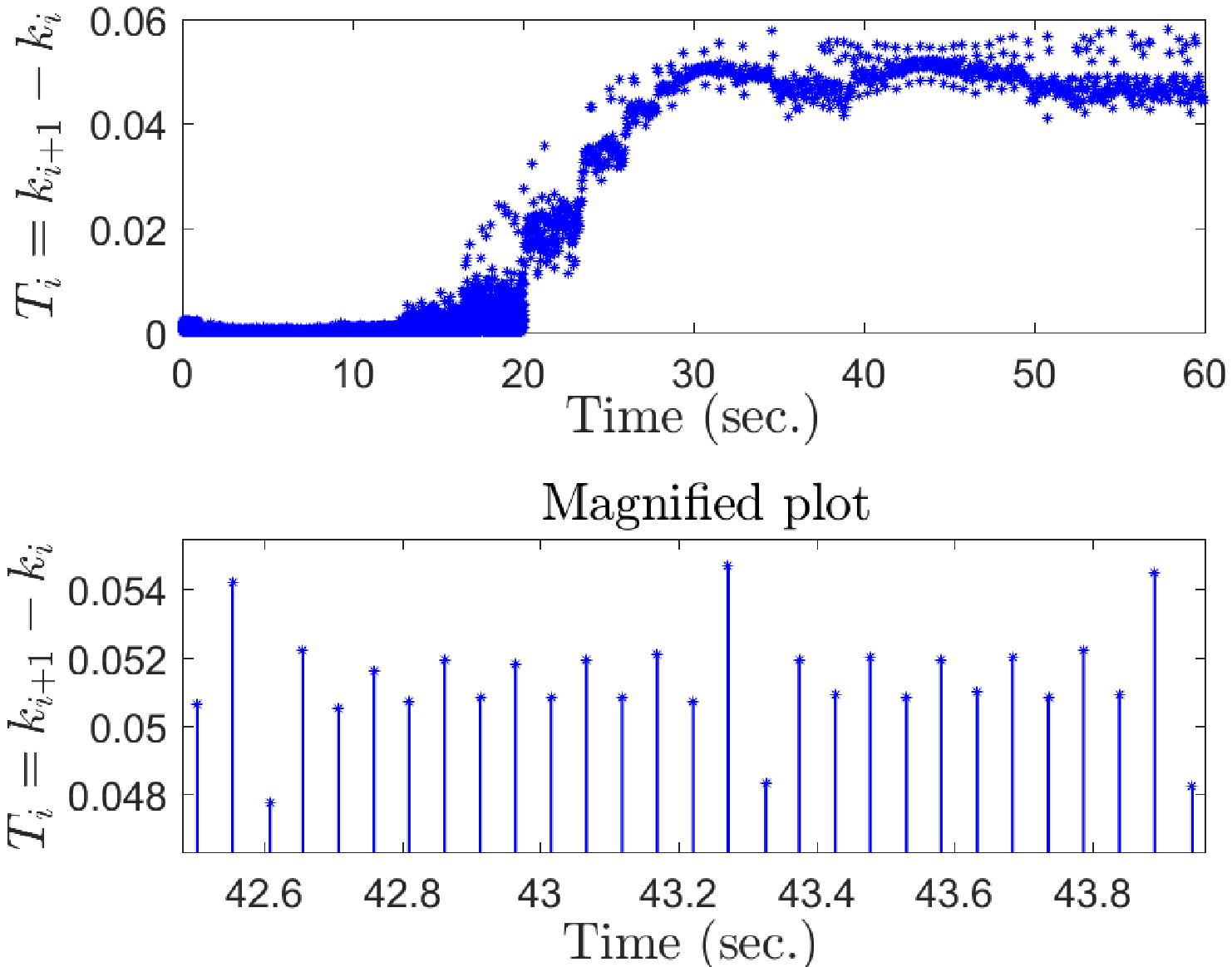}}
\small\caption{Evolution of inter-event time}
\label{fig4}
\end{figure}

\quad The plots for this example are shown in Figs. \ref{fig1}-\ref{fig4}. The evolution of state trajectories is shown in Fig. \ref{fig1}, and it shows that the system trajectories are practically bounded under the application of event-triggered control. Fig. \ref{fig2} shows the evolution of the sliding variable and it is observed that the sliding trajectory reaches the practical QSMB $\Omega_1=0.0286$ in the finite time $t = 2.6$sec. Fig. \ref{fig3} shows the event-triggered SMC. The inter-event time $T_i=k_{i+1}-k_i$ is shown in Fig. \ref{fig4}, and it is evident from the figure that control input is kept constant for long time periods. $T_i=0.0581$ sec is the maximum time gap between two consecutive control updating instants, which is a big multiple of sampling interval i.e., $581\tau$. For a run time of $60$sec, the total number of control updating instants are $6\times 10^{5}$ for time-triggered control and $90475$ for event-triggered control strategy. In the case of time-triggered control, more resources and control effort is required as control is updated after every $\tau$ period. While on the other hand, in the event-triggering framework, control is updated aperiodically which results in a constant control signal for long time periods and it reduces the resource utilization significantly.
\begin{figure}[h]
\centering{\includegraphics[width=1\linewidth]{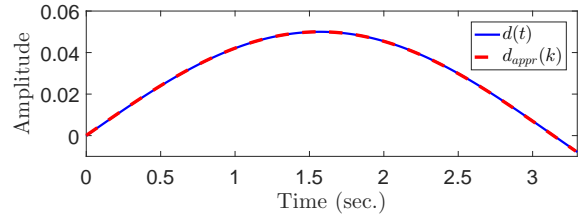}}
\small\caption{Plot of disturbance}
\label{fig5}
\end{figure}

\quad The delta operator system \eqref{delexam} has relative degree one i.e., $CB_{\delta \tau} \neq 0$ but it is a non-minimum phase system. Hence the method presented in \cite{shimd} for computing an equivalent matched and bounded disturbance cannot be applied here. However, we can use another candidate of disturbance $d_{\text{appr}}(k)$ which is an approximate version of $d(k)$ as given in \cite{shimd}. Thus without using Assumption \ref{assum3} on the disturbance and computing $d_{\text{appr}}(k)$ using the method presented in \cite{shimd}, the delta operator system can be represented as 
\begin{align}
\label{newsys}
\delta\xi(k) &= A_{\delta \tau}\xi(k)+B_{\delta \tau}(u(k)+d_{\text{appr}}(k)).
\end{align}
The approximate sampled disturbance for the above system is obtained as $d_{\text{appr}}(k) = \dfrac{1}{h_0}Cw(k)$, where $h_0 = C B_{\delta \tau}$ and $w(k) = \dfrac{1}{\tau} \int\limits_{k\tau}^{(k+1)\tau}e^{A((k+1)\tau-s)}Bd(s)\mathrm{d}s $. The plot of approximate matched disturbance is given in Fig. \ref{fig5}. So without using Assumption \ref{assum3}, we can apply the proposed method to the system \eqref{newsys} using the upper bound of the disturbance $d_{\text{appr}}(k)$. The system behaviour will be same as shown in Figs. \ref{fig1}-\ref{fig4}.
\vspace{-0.05in}

\subsection{Example 2}
\vspace{-0.10in}

Consider a numerical example given by \eqref{sys1} with 
\vspace{-0.34in}

\begin{align*}
A &= \begin{bmatrix}
-2 & 1 & 0 \\  -2 & 0 & 1 \\  -1 & 0 & 0
\end{bmatrix}, \ B = \begin{bmatrix}
0 \\ 5 \\ 4
\end{bmatrix}, \ \mbox{and} \ C = \begin{bmatrix}
1 & 0 & 0
\end{bmatrix}.
\end{align*}
\vspace{-0.32in}

The disturbance is taken as $d(t) = 0.05\sin(2t)$. Here we show simulation results for the proposed control strategy without using Assumption \ref{assum3} on the disturbance. For this, the method presented in \cite{shimd} to obtain an equivalent matched sampled disturbance for the discrete-time system is modified, so that it can be used for delta operator system. To apply this method the delta operator system should be of relative degree one i.e., $CB_{\delta \tau} \neq 0$ and minimum phase. The delta operator system for sampling period $\tau = 10^{-2}$sec is given by
\vspace{-0.34in}

\begin{align*}
\delta\xi(k) = A_{\delta \tau}\xi(k)+B_{\delta \tau}(u(k)+\hat{d}(k))
\end{align*}
\vspace{-0.34in}

and it satisfies both the properties. The disturbance $\hat{d}(k)$ can be obtained as
\vspace{-0.34in}

\begin{align*}
\hat{d}(k) &= \mathcal{C}_m \bar{\zeta}(k)+C\dfrac{1}{h_0}w(k), \\
\bar{\zeta}(k+1) &= \mathcal{A}_m \bar{\zeta}(k) +\dfrac{1}{h_0}w(k), \quad \bar{\zeta}(0)=0, 
\end{align*}
\vspace{-0.34in}

where $\mathcal{A}_m = \hat{A}-\dfrac{1}{h_0}\hat{A}B_{\delta \tau}C$, $\mathcal{C}_m=C\mathcal{A}_m$, $\hat{A} := A_{\delta \tau}\tau +I$ and $w(k) = \dfrac{1}{\tau} \int\limits_{k\tau}^{(k+1)\tau}e^{A((k+1)\tau-s)}Bd(s)\mathrm{d}s $. The derivation of above given expression of equivalent matched and bounded disturbance is followed from \cite{shimd}. The system matrix and input matrix for $\tau = 10^{-2}$ sampled delta operator system are 
\vspace{-0.35in}

\begin{align*}
A_{\delta \tau} &= \begin{bmatrix}
 -1.9900 & 0.9900 & 0.0050 \\ -1.9850 & -0.0100 & 1.0000 \\ -0.9900 & -0.0050 & -0.0000
\end{bmatrix}, \ B_{\delta \tau} = \begin{bmatrix}
0.0249 \\ 5.0198 \\ 3.9999
\end{bmatrix}.
\end{align*}
\vspace{-0.25in}

\begin{figure}[h]
\centering{\includegraphics[width=1\linewidth]{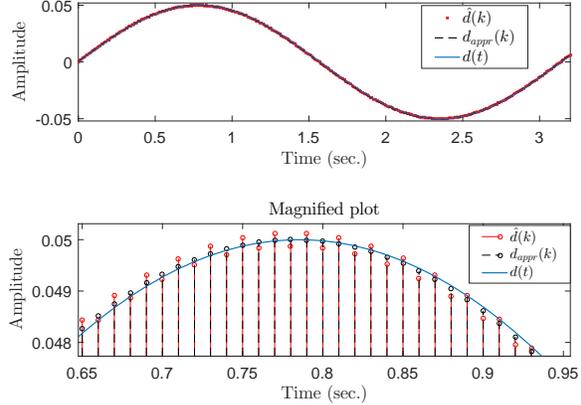}}
\small\caption{Plot of disturbance}
\label{fig6}
\end{figure}
\begin{figure}[h]
\centering{\includegraphics[width=1\linewidth]{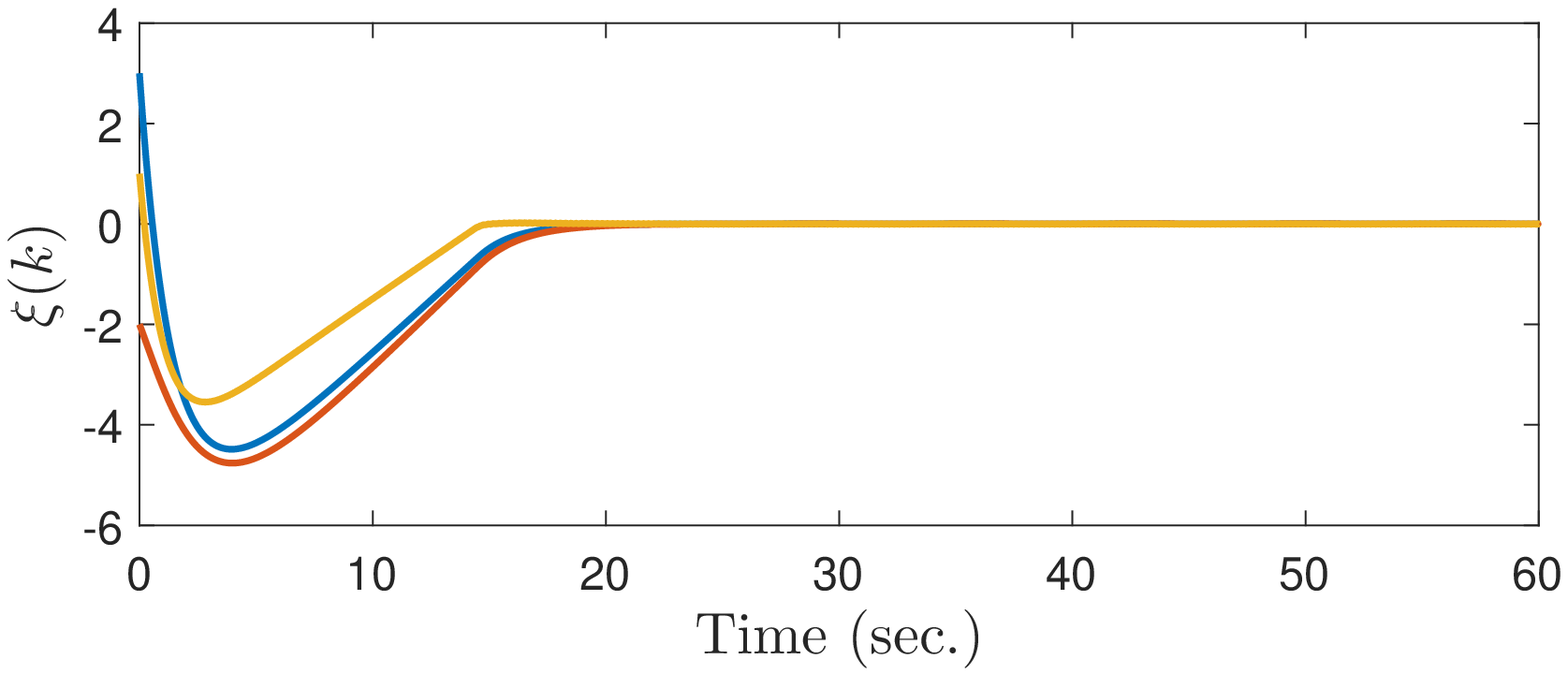}}
\small\caption{State trajectories}
\label{fig7}
\end{figure}
\begin{figure}[h]
\centering{\includegraphics[width=1\linewidth]{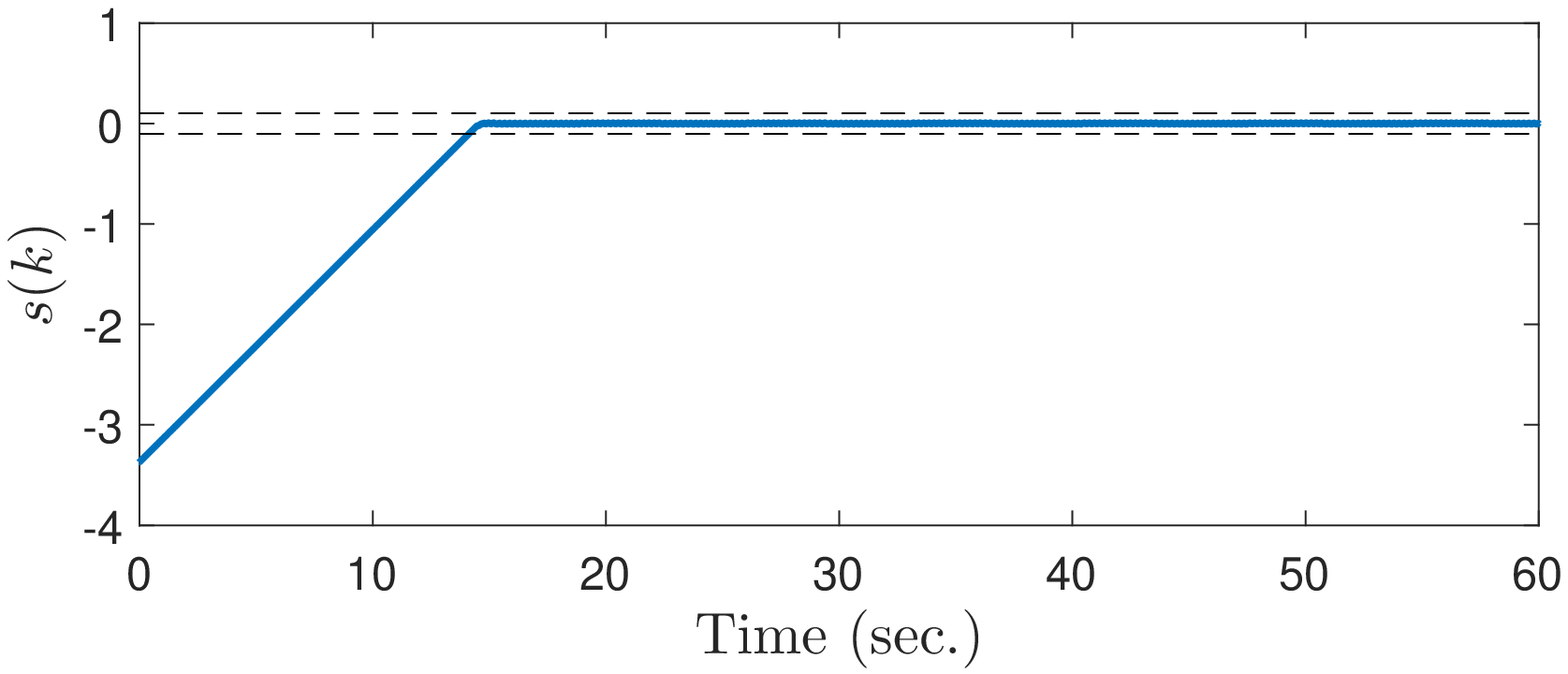}}
\small\caption{Sliding trajectory}
\label{fig8}
\end{figure}
\begin{figure}[h]
\centering{\includegraphics[width=1\linewidth]{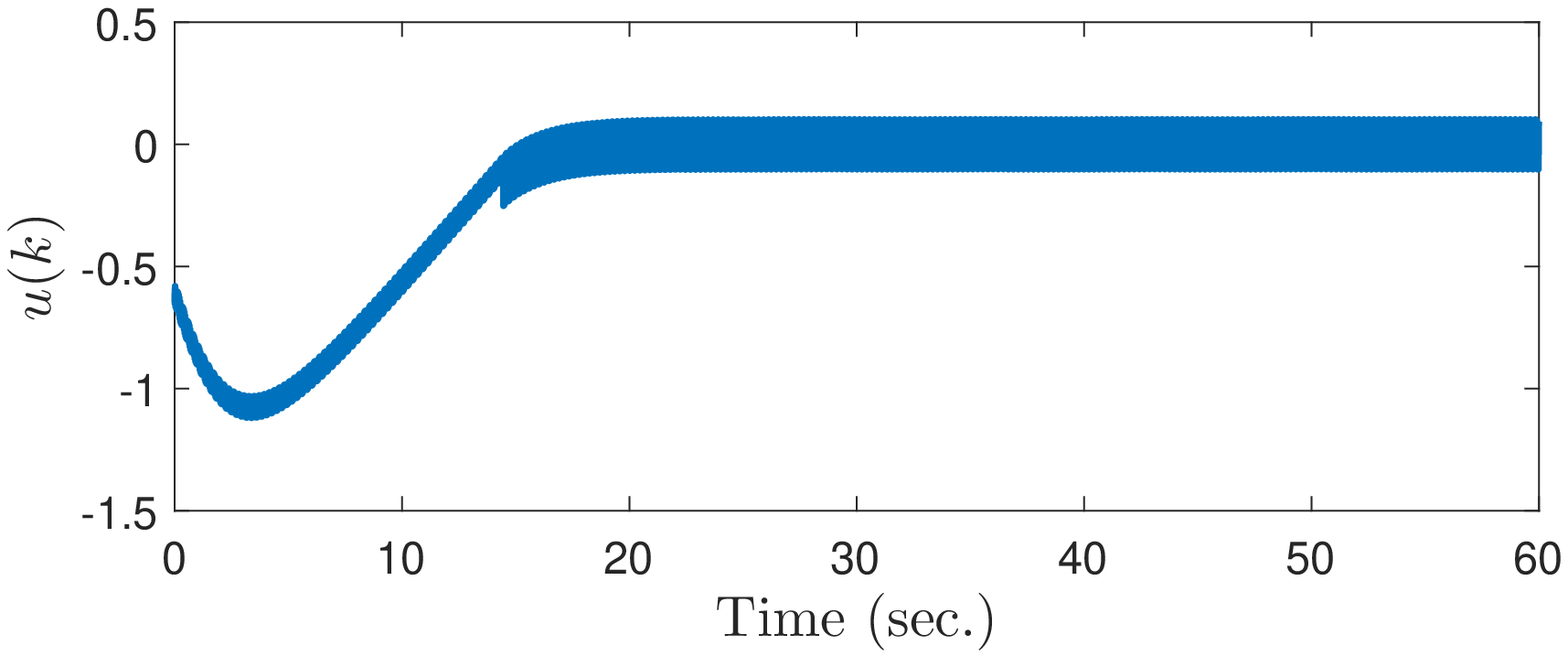}}
\small\caption{Control input}
\label{fig9}
\end{figure}
\begin{figure}[h]
\centering{\includegraphics[width=1\linewidth]{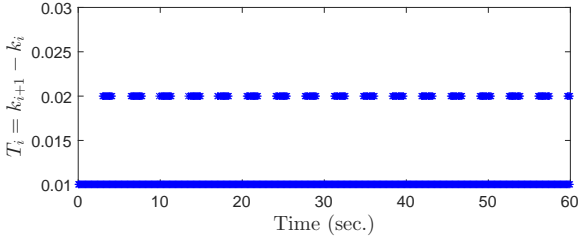}}
\small\caption{Evolution of inter-event time}
\label{fig10}
\end{figure}
\quad The initial condition is taken as $\xi_0=[3 \ -2 \ 1]^\top$ and $c = [-1.024 \ 0.6836 \ 1]^\top$. Other parameters are chosen as $N = 3$, $\alpha = 0.2$, $\sigma = 0.9$ and $\epsilon=0.2412$. 
\vspace{-0.05in}

\quad The plot of equivalent sampled disturbance $\hat{d}(k)$ along with $d_{\text{appr}}(k)$ and $d(t)$ is given in Fig. \ref{fig6}, and it can be noted that the upper bound on $\hat{d}(k)$ is larger than the upper bound on continuous disturbance $d(t)$. The maximum value of $\hat{d}(k)$ is obtained as $0.0501$. Thus we can take the upper bound on disturbance as $d_0 = 0.0502$. Using this upper bound, the proposed event-triggered control strategy is applied to the delta operator system. Fig. \ref{fig7} shows state trajectories and Fig. \ref{fig8} shows the sliding trajectory. The value of practical QSMB is $0.1119$. The event-triggered SMC is shown in Fig. \ref{fig9} and inter-event time in Fig. \ref{fig10}. The remaining explanations of the figures are the same as given in Example $1$.
\vspace{-0.05in}
  
\section{Conclusion} \label{con}
\vspace{-0.10in}

In this paper, an alternative formula for MRSE is proposed which is used for designing output feedback based control for the delta operator system. The proposed formula enhances the numerical accuracy for small sampling periods. To avoid the numerical singularity for small sampling periods, a new observability matrix is proposed which recovers the continuous-time counterpart as the sampling period tends to zero. SMC law is designed for the delta operator system using MRSE. When the sampling period is very small, the frequency of control updates is very high. Thus to reduce the number of control updates, an event-triggered SMC is proposed for the delta operator system using MRSE results. In the event-triggering framework, the band size increases, but control updates are reduced considerably. Moreover, a good trade-off between the triggering instants and the band size has to be maintained to achieve the desired performance.
\vspace{-0.08in}

\appendix
\section{Proof of Theorem \ref{th2}}\label{app1}
\vspace{-0.10in}

Consider the Lyapunov function $V(s(k))=s^2(k).$ Applying delta operator on the Lyapunov function gives 
\vspace{-0.36in}

\begin{align}
\label{lyp}
\delta V(s(k)) = \delta s(k)\left(\tau \delta s(k) +2s(k)\right).
\end{align}
\vspace{-0.36in}

Now, using modified reaching law \eqref{eq6} in the above relation results in
\vspace{-0.36in}

\begin{align*}
\delta V(s(k))&=\tau\big(\tilde{d}^2(k)+\epsilon^2+f^2(k-1)-2\epsilon\tilde{d}(k)\operatorname{sgn}\bar{s}(k)  \\
&\quad +2\tilde{d}(k)f(k-1) -2\epsilon f(k-1)\operatorname{sgn}\bar{s}(k) \big) \\
&\quad +2s(k)\tilde{d}(k)+2s(k)f(k-1)-2\epsilon s(k)\operatorname{sgn}\bar{s}(k).
\end{align*}
\vspace{-0.36in}

We know that $\operatorname{sgn} \bar{s}(k)= \operatorname{sgn} s(k)$ for $|s(k)|>l_m$. Thus
\vspace{-0.36in}

\begin{align*}
\delta V(s(k)) & = \tau\big(\tilde{d}^2(k)+\epsilon^2+f^2(k-1)-2\epsilon\tilde{d}(k)\operatorname{sgn}s(k)\\
            & \quad+2\tilde{d}(k)f(k-1) -2\epsilon f(k-1)\operatorname{sgn}s(k) \big) \\
            & \quad +2s(k)\tilde{d}(k)+2s(k)f(k-1)-2\epsilon s(k)\operatorname{sgn}s(k) \\
            & \leq \tau\big(\lvert \tilde{d}(k) \rvert^2+\epsilon^2+\lvert f(k-1)\rvert^2+2\epsilon \lvert f(k-1)\rvert  \\
            & \quad  +2\epsilon \lvert \tilde{d}(k)\rvert + 2 \lvert \tilde{d}(k)\rvert \lvert f(k-1)\rvert \big) \\ & \quad +2\lvert s(k)\rvert (\lvert \tilde{d}(k)\rvert + \lvert f(k-1)\rvert -\epsilon) .
\end{align*}
\vspace{-0.36in}

Using $\lvert{\tilde{d}(k)}\rvert\leq d_m$, $\lvert l(k)\rvert \leq l_m$ and $\lvert{f(k)}\rvert\leq f_m$ gives
\vspace{-0.36in}

\begin{align*}
\delta V(s(k)) &\leq \tau \left( d_{m}^{2} + \epsilon^2+ f_{m}^{2} + 2d_m\epsilon+ 2d_mf_m +2f_m\epsilon \right) \\
            & \quad +2\left( d_m+f_m-\epsilon \right)\lvert s(k)\rvert \\
            &= -2\left( \epsilon-d_m-f_m \right) \left( \lvert s(k)\rvert -\frac{\tau(\epsilon+d_m+f_m)^2}{2(\epsilon-d_m-f_m)} \right).
\end{align*}
\vspace{-0.32in}

Having $\epsilon>d_m+f_m$, we define $a_1 = 2\left(\epsilon-d_m-f_m\right)$ and $b_1 = \tau\left(\epsilon+d_m+f_m\right)^2/ a_1$. Now write the last inequality as 
\vspace{-0.36in}

\begin{align}
\label{lyaineq}
\delta V(s(k)) \leq -a_1 (\left \lvert s(k)\rvert -b_1 \right).
\end{align}
\vspace{-0.36in}

It follows from \eqref{lyaineq} that whenever $|s(k)|> \text{max} \left\{ b_1,l_m \right\}$, we get $\delta V(s(k)) < 0$. This ensures the finite time reachability of the sliding variable. If either of the conditions $|s(k)|>b_1$ or $|s(k)|>l_m$ is not satisfied, we cannot guarantee that $\delta V(s(k)) < 0$, but we can say that the trajectory gets confined in a band around the sliding manifold. Thus we calculate QSMB using the fact that $|s(k)|<b_1$ and $|s(k)|< l_m$ do not result in decreasing Lyapunov function. We divide the proof in two cases: (i) $l_m > b_1$, (ii) $l_m < b_1$.
\vspace{-0.09in}

\quad (i) Suppose $|s(k)|\leq l_m$, then we calculate the value of sliding variable at next time instant using reaching law \eqref{eq6} as  
\vspace{-0.38in}

\begin{align*}
|s(k+1)| &= |s(k)-\tau\epsilon \operatorname{sgn}\bar{s}(k)+\tau\tilde{d}(k)+\tau f(k-1)| \\
& \leq l_m+\tau(\epsilon+d_m+f_m).
\end{align*}
\vspace{-0.37in}

Since $l_m+\tau(\epsilon+d_m+f_m)>l_m$, $\delta V(s(k+1)) < 0$ satisfies. Thus in this case we obtain the maximum deviation in sliding trajectory in one time step as $l_m+\tau(\epsilon+d_m+f_m)$.
\vspace{-0.09in}

\quad (ii) Suppose $l_m < |s(k)| \leq b_1$, then we can obtain \eqref{lyaineq} using the fact that $\operatorname{sgn} \bar{s}(k)= \operatorname{sgn} s(k)$ for $|s(k)|>l_m$. But $\delta V(s(k))\nless 0$ as $|s(k)| \leq b_1$, so we find $s(k+1)$ using \eqref{lyaineq} as 
\vspace{-0.36in}

\begin{align*}
s^2(k+1) & \leq s^2(k) +\tau a_1 b_1-  \tau a_1\lvert s(k) \rvert  < s^2(k) + \tau a_1b_1.
\end{align*}
\vspace{-0.36in}

Hence we have $|s(k+1)| \leq (b_1^2+\tau a_1 b_1)^{1/2}$ and $(b_1^2+\tau a_1 b_1)^{1/2}>b_1$, so $\delta V(s(k+1)) < 0$. Now if $|s(k)| \leq l_m$, using reaching law we have $|s(k+1)|<l_m+\tau(\epsilon+d_m+f_m)$. Therefore, in this case, the maximum deviation in sliding trajectory in one time step will be maximum of $l_m+\tau(\epsilon+d_m+f_m)$ and $(b_1^2+\tau a_1 b_1)^{1/2}.$
\vspace{-0.09in}

\quad
For all $|s(k)|\leq \max \left\{ b_1,l_m \right\}$, $|s(k+1)|$ lies in the set $ \left\lbrace x \in \mathbb{R}^n : \lvert c^\top x \rvert \leq \max \left\{l_m+\tau(\epsilon+d_m+f_m),(b_1^2+\tau a_1 b_1)^{1/2} \right\} \right\rbrace $ and this set is invariant because of above arguments. As a result, the QSMB is obtained as
\vspace{-0.36in}

\begin{equation}
\label{band2}
\Omega =  \max \left\{ l_m+\tau(\epsilon+d_m+f_m), (b_1^2+\tau a_1 b_1)^{1/2} \right\}.
\end{equation}
\vspace{-0.36in}

This completes the proof. \hfill \qed
\vspace{-0.05in}

\section{Proof of Lemma \ref{le1}}\label{app3}
\vspace{-0.10in}

Using \eqref{conerr}, we can write
\vspace{-0.3in}

\begin{small}
\begin{align}
\label{errdy1} \begin{bmatrix}
x(k) \\ e_u(k)
\end{bmatrix} =& \begin{bmatrix}
I & 0 \\ -F & I
\end{bmatrix}\begin{bmatrix}
x(k) \\ u(k)
\end{bmatrix}+\begin{bmatrix}
0 \\ FTL_{u}^{\delta}d(k-1) -g_1(k)
\end{bmatrix} 
\end{align}
\begin{align}
\label{errdy} \begin{bmatrix}
x(k+1) \\ e_u(k+1)
\end{bmatrix} =& \begin{bmatrix}
I & 0 \\ -F & I
\end{bmatrix}\begin{bmatrix}
x(k+1) \\ u(k+1)
\end{bmatrix}+\begin{bmatrix}
0 \\ FTL_{u}^{\delta}d(k) -g_1(k+1)
\end{bmatrix}.
\end{align}
\end{small}
\vspace{-0.20in}

From the system dynamics in \eqref{trsys1}, the relation $x(k) = \bar{x}(k)+ TL_{u}^{\delta}d(k-1)$, and \eqref{con1} we obtain
\vspace{-0.32in}

\begin{align*}
u(k+1)   
		&= F \left(\bar{A}x(k) +\bar{B} u(k) +\bar{B} d(k) 
		 -TL_{u}^{\delta}d(k)\right) +g_1(k+1).
\end{align*}
\vspace{-0.32in}

Now from above equation we get
\vspace{-0.28in}

\begin{small}
\begin{align*}
\nonumber \begin{bmatrix}
x(k+1) \\ u(k+1)
\end{bmatrix} =& \begin{bmatrix}
\bar{A}  & \bar{B} \\ F\bar{A}  & F\bar{B}
\end{bmatrix} \begin{bmatrix}
x(k) \\ u(k)
\end{bmatrix} + \begin{bmatrix}
\bar{B}d(k) \\ F\left(\bar{B}-T L_{u}^{\delta}\right)d(k)+g_1(k+1)
\end{bmatrix}.
\end{align*}
\end{small}
\vspace{-0.22in}

Using the above equation in \eqref{errdy} results in
\vspace{-0.34in}

\begin{align*}
\nonumber \begin{bmatrix}
x(k+1) \\ e_u(k+1)
\end{bmatrix} =& \begin{bmatrix}
\bar{A}  & \bar{B} \\ 0 & 0
\end{bmatrix} \begin{bmatrix}
x(k) \\ u(k)
\end{bmatrix}+ \begin{bmatrix}
\bar{B} \\  0
\end{bmatrix} d(k).
\end{align*}
\vspace{-0.28in}

Substituting $\begin{bmatrix}
x(k) \\ u(k)
\end{bmatrix}$ from \eqref{errdy1} in above equation yields
\vspace{-0.28in}

\begin{align*}
\begin{bmatrix}
x(k+1) \\ e_u(k+1)
\end{bmatrix} 
=& \begin{bmatrix}
\bar{A} + \bar{B} F  & \bar{B}  \\ 0 & 0
\end{bmatrix} \begin{bmatrix}
x(k) \\ e_u(k)
\end{bmatrix}  -\begin{bmatrix}
 g_1(k) \\ 0
\end{bmatrix} 
 \\ & + \begin{bmatrix}
\bar{B} \left( FTL_{u}^{\delta}d(k-1) + d(k) \right) \\ 0
\end{bmatrix}.
\end{align*}
\vspace{-0.28in}

It is evident from the above relation that error in control becomes zero in one time step so there is no propagation of error in the system due to initial unavailability of the state.  \hfill \qed
\vspace{-0.05in}

\section{Proof of Theorem \ref{th3}}\label{app2}
\vspace{-0.10in}

Applying the control law \eqref{contrig} to the dynamics of sliding variable \eqref{eq8} gives
\begin{align*}
\nonumber \delta s(k) =& c^\top \bar{A} x(k)-c^\top \bar{A} \bar{x}(k_i)-\epsilon\operatorname{sgn}\bar{s}(k_i)+\tilde{d}(k) \\
\nonumber             =& c^\top \bar{A} \bar{x}(k)+f(k-1)-c^\top  \bar{A} \bar{x}(k_i)
                        -\epsilon\operatorname{sgn}\bar{s}(k_i)+\tilde{d}(k)\\
                      =& -c^\top \bar{A} \bar{e}(k)-\epsilon\operatorname{sgn}\bar{s}(k_i)+\tilde{d}(k)+f(k-1).
\end{align*}
\vspace{-0.36in}

Substituting above in the Lyapunov function \eqref{lyp} yields
\vspace{-0.05in}
\begin{align*}
\delta V(s(k))& =\tau\big((c^\top\bar{A} \bar{e}(k))^2 +2c^\top\bar{A} \bar{e}(k)\epsilon\operatorname{sgn}\bar{s}(k_i) +\epsilon^2 \\
           & \quad +(\tilde{d}(k)+f(k-1))^2 -2c^\top \bar{A} \bar{e}(k)\tilde{d}(k)\\
           & \quad -2c^\top \bar{A} \bar{e}(k)f(k-1)-2\epsilon\operatorname{sgn}\bar{s}(k_i)\tilde{d}(k)\\
           & \quad -2\epsilon\operatorname{sgn}\bar{s}(k_i)f(k-1)\big) -2s(k)\epsilon\operatorname{sgn}\bar{s}(k_i)\\
           & \quad -2s(k)c^\top \bar{A} \bar{e}(k)+2s(k)(\tilde{d}(k)+f(k-1)).
\end{align*}
As we have discussed earlier, there is a band around the sliding manifold because of event-triggering strategy, so it is to be noted that the sign of $\bar{s}(k)$ does not change until the sliding manifold is reached, and as a result $\operatorname{sgn}\bar{s}(k_i)=\operatorname{sgn}\bar{s}(k)$. Moreover, as mentioned earlier, $\operatorname{sgn}\bar{s}(k)=\operatorname{sgn}s(k)$, if $|s(k)| >l_m$. Using this in the above relation results in
\vspace{-0.05in}
\begin{align*}
\delta V(s(k)) & \leq \tau \big(\lvert c^\top \bar{A} \bar{e}(k)\rvert^2+\lvert\tilde{d}(k)+f(k-1)\rvert^2+\epsilon^2  \\
            & \quad +2\epsilon\lvert c^\top \bar{A} \bar{e}(k)\rvert +2\epsilon\lvert \tilde{d}(k)+f(k-1)\rvert \\
            & \quad +2\lvert c^\top \bar{A} e(k)\rvert (\lvert\tilde{d}(k)+f(k-1)\rvert )\big)-2\epsilon\lvert s(k)\rvert\\
            & \quad   +2\left| s(k)\right| \lvert c^\top \bar{A} \bar{e}(k)\rvert+2 \lvert s(k)\rvert \lvert\tilde{d}(k)+f(k-1)\rvert.
\end{align*}
\vspace{-0.34in}

Using $\lvert{\tilde{d}(k)}\rvert\leq d_m$, $\lvert f(k) \rvert \leq f_m$, $\lvert l(k) \rvert \leq l_m$ and condition \eqref{eq10} into the above equation gives
\vspace{-0.34in}

\begin{align}
\nonumber \delta V(s(k)) & \leq \tau (\alpha^2+(d_{m}+f_m)^{2}+\epsilon^2+2\alpha\epsilon+2\epsilon(d_m+f_m)\\
\nonumber    &\quad +2\alpha (d_{m}+f_m)) +2(\alpha+d_m+f_m-\epsilon)\lvert s(k)\rvert.
\end{align}
\vspace{-0.34in}

Having $\epsilon>\alpha+d_m+f_m$, we define $a_3 = 2\left(\epsilon-\alpha-d_m-f_m\right)$ and $b_3 = \tau\left(\epsilon+\alpha+d_m+f_m\right)^2/ a_3$. Now we can write the last inequality as 
\vspace{-0.36in}

\begin{align}
\label{lastineq} \delta V(s(k))  & =-a_3 \left(\lvert s(k)\rvert- b_3 \right).
\end{align}
\vspace{-0.36in}

From \eqref{lastineq}, it can be noticed that when $\left| s(k)\right| > b_3 $ we can achieve $\delta V(k) < 0$. Now following the explanation used in the proof of Theorem \ref{th2}, it can be seen that the set $\lbrace x \in \mathbb{R}^n : \lvert c^\top x \rvert\leq \theta \rbrace $, where $\theta = \max \left\{l_m+\tau(\epsilon+d_m+f_m),(b_3^2+\tau a_3 b_3)^{1/2} \right\} $, is an invariant set. However, in the band $\left| s(k)\right| \leq \theta$, whether $\operatorname{sgn}\bar{s}(k_i)=\operatorname{sgn}\bar{s}(k)$ holds or not, cannot be told and as a result, \eqref{lastineq} cannot be achieved. This is because when $\bar{s}(k)$ changes sign, $\bar{s}(k_i)$ may or may not change sign, but once the next triggering instant comes, $\bar{s}(k_i)$ will also change sign. We calculate the maximum deviation of $\bar{s}(k)$ in one triggering instant as
\vspace{-0.34in}

\begin{align*}
\left| c^\top \bar{x}(k)-c^\top \bar{x}(k_i) \right| =\left| c^\top \bar{e}(k) \right| & \leq \|c\| \|\bar{e}(k)\| < \alpha \| \bar{A} \|^{-1}.
\end{align*}
\vspace{-0.34in}

The above inequality is obtained using relation \eqref{eq10}. If the trajectory crosses the sliding manifold before the occurrence of the next triggering instant, then the trajectory remains bounded in the region $\left\lbrace \bar{x} \in \mathbb{R}^n : \lvert c^\top \bar{x} \rvert\leq \alpha \| \bar{A} \|^{-1} \right\rbrace $.  Now using \eqref{sbar} and $|\bar{s}(k)| \leq \alpha \| \bar{A} \|^{-1}$, we calculate $|s(k)| \leq \alpha \| \bar{A} \|^{-1} +l_m$. If $\alpha \| \bar{A} \|^{-1}+l_m $ is larger than $\theta$, the trajectory still remain bounded in the region $\left\lbrace x \in \mathbb{R}^n : \lvert c^\top x \rvert\leq \alpha \| \bar{A} \|^{-1} +l_m \right\rbrace $. This is because when the trajectory crosses value $ \theta $, no triggering instant is generated, and hence no control is updated. One more thing to be noted is that the value of $\alpha \| \bar{A} \|^{-1}$ remains constant for a fixed $\alpha$, as it is independent of the sampling period. However, the value of $\theta$ varies with $\alpha$ and $\tau$ as it depends on both. So if $\theta$ is larger than $\alpha \| \bar{A} \|^{-1}+l_m$ the trajectory would be bounded in the region $\left\lbrace x \in \mathbb{R}^n : \lvert c^\top x \rvert\leq \theta \right\rbrace $. Hence the bound of sliding variable is the maximum of $\theta$ and $\alpha \| \bar{A} \|^{-1}+l_m$. Therefore, the practical QSMB is given by
\begin{equation}
\label{band4}
\Omega_1 = \max \left\lbrace l_m+\tau(\epsilon+d_m+f_m), (b_3^2+\tau a_3 b_3)^{1/2}, \alpha \| \bar{A} \|^{-1}+l_m  \right\rbrace.
\end{equation}
This proves that trajectory is driven into practical QSMB in finite time $\hat{k}$ and remain there for all $k \geq \hat{k}$. This ensures the existence of practical quasi-sliding mode in the system. \hfill \qed

\end{document}